\documentclass[
    ,final            
  ]
  {aipproc}
\layoutstyle{8x11double}

\begin{document}

\title{Quark masses and strong CP violation
}

\classification{11.30.Er, 12.39.Fe, 11.15.Ha, 11.10.Gh}
\keywords      {QCD, quark masses, strong CP violation}

\author{Michael Creutz
\footnote
{I thank the Alexander von Humboldt Foundation for supporting visits
 to the University of Mainz.  This manuscript has been authored under
 contract number DE-AC02-98CH10886 with the U.S.~Department of Energy.
 Accordingly, the U.S. Government retains a non-exclusive,
 royalty-free license to publish or reproduce the published form of
 this contribution, or allow others to do so, for U.S.~Government
 purposes.}  }
{ address={ Physics Department, Brookhaven National
 Laboratory, Upton, NY 11973, USA }}

\begin{abstract}
{ Two flavor QCD involves three independent mass parameters for which
non-perturbative effects are not universal.  This precludes matching
lattice and perturbative results for non-degenerate quarks and
eliminates a vanishing up quark mass as a viable solution to the
strong CP problem.  }
\end{abstract}
\maketitle

\begin{figure}
\centering
\label{scattering}
\includegraphics[width=1.7in]{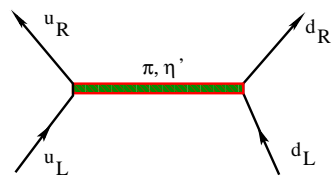}
\caption{Both pion and eta prime exchange contribute towards spin flip
  scattering between up and down quarks. Because these mesons are
  non-degenerate, this scattering is not helicity suppressed.}
\end{figure}

In massless two-flavor QCD, chiral symmetry breaking gives rise to
three massless Goldstone pions.  In contrast, the two flavor analog of
the eta prime meson acquires a mass from the anomaly.  Thus, as
shown schematically in Fig.~\ref{scattering}, meson exchange will
contribute to a hypothetical quark spin-flip scattering experiment.

Now turn on a small $d$ quark mass. This allows connecting the
ingoing and outgoing $d$ quark lines in Fig.~\ref{scattering}, and
gives a mixing between the left and right handed $u$ quark.  The
presence of a non-zero $d$ quark mass creates an effective mass for
the $u$ quark, even if the latter initially vanishes.
Non-perturbative effects renormalize $m_u/ m_d$.  If this ratio is
zero at some scale, it cannot remain so for all scales.  This cross
talk between the masses of different quark species has been noted
several times in the past \cite{Georgi:1981be} and contradicts the
lore that mass renormalization is flavor blind.  The practice of
matching lattice calculations to $\overline{MS}$ is problematic when
$m_u\ne m_d$.

A general mass term is an electrically neutral quadratic form that
transforms as a Lorentz singlet.  This leaves four candidates $
m_1\overline\psi\psi+ m_2\overline\psi\tau_3\psi+
im_3\overline\psi\gamma_5\psi+ im_4\overline\psi\gamma_5\tau_3\psi.  $
The massless limit should have the flavored chiral symmetry under
$\psi \longrightarrow e^{i\gamma_5 \tau_\alpha\phi_\alpha}\psi$.  With
the masses present, this mixes $m_1$ with $m_4$ and $m_2$ with $m_3$.
The four mass terms are not independent and one can select any one of
the $m_i$ to vanish and a second to be positive.  The chiral anomaly
is responsible for the singlet rotation $
\psi \longrightarrow e^{i\gamma_5 \phi}\psi$
not being a valid symmetry \cite{Fujikawa:1979ay}.  This rotation
does, however, allow one to remove any topological term from the gauge
part of the action.  Assume this has been done.

Adopt the common choice $m_4=0$ and $m_1$ as the average quark
mass. Then $m_2$ is the quark mass difference and $m_3$ is CP
violating.  The possible presence of $m_3$ represents the strong CP
problem.

Strong interactions preserve CP to high accuracy.  With the above
conventions, it is natural to ask why is $m_3$ so small?  One proposed
solution is that the up quark mass might vanish, allowing a flavored
chiral rotation to remove any phases from the quark mass matrix.

Why is this not a sensible approach?  From the above, one can define
the up quark mass as $ m_u\equiv m_1+m_2+im_3.$ But the quantities
$\{m_1,m_2,m_3\}$ are independent parameters with different symmetry
properties.  As discussed earlier, the combination $m_1+m_2=0$ is
scale and scheme dependent.  While it may be true that $
m_1+m_2+im_3=0$ implies $m_3=0,$ this would depend on scale and
should be regarded as ``not even wrong.''

\end{document}